\documentclass[runningheads]{llncs}
\usepackage[T1]{fontenc}
\usepackage{amsmath,amssymb,amsfonts}
\usepackage{bm}
\usepackage{graphicx}
\usepackage{subcaption}
\usepackage{xcolor}
\usepackage[nocompress]{cite}

\newcommand{\Ra}{\text{Ra}}
\newcommand{\Nu}{\text{Nu}}

\begin{document}
\title{Control of Rayleigh-Bénard Convection: Effectiveness of Reinforcement Learning in the Turbulent Regime}
% Michiel: alternative: Control of Rayleigh-Bénard Convection: Effectiveness of Reinforcement Learning under Increasing Turbulence.
% Michiel: alternative: On the Effectiveness of Reinforcement Learning for Controlling Turbulent Rayleigh-Bénard Convection
%
\titlerunning{Control of Rayleigh-Bénard Convection}

\author{Thorben Markmann\inst{1}\orcidID{0000-0003-3145-4577} \and
Michiel Straat\inst{1}\orcidID{0000-0002-3832-978X} \and
Sebastian Peitz\inst{2}\orcidID{0000-0002-3389-793X} \and
Barbara Hammer\inst{1}\orcidID{0000-0002-0935-5591}}
%\author{Anonymous submission}

\authorrunning{T. Markmann et al.}
%\authorrunning{Anonymous submission}

\institute{Bielefeld University, Center for Cognitive Interaction Technology (CITEC), Inspiration 1, 33619 Bielefeld, Germany\\\email{\{tmarkmann,mstraat,bhammer\}@techfak.uni-bielefeld.de} \and Department of Computer Science \& Lamarr Institute for Machine Learning and Artificial Intelligence, TU Dortmund University, Joseph-von-Fraunhofer-Straße 25, 44227 Dortmund, Germany\\\email{sebastian.peitz@tu-dortmund.de}}
\maketitle              % typeset the header of the contribution

\begin{abstract}
Data-driven flow control has significant potential for industry, energy systems, and climate science.
In this work, we study the effectiveness of Reinforcement Learning (RL) for reducing convective heat transfer in the 2D Rayleigh-Bénard Convection (RBC) system under increasing turbulence.
We investigate the generalizability of control across varying initial conditions and turbulence levels and introduce a reward shaping technique to accelerate the training.
RL agents trained via single-agent Proximal Policy Optimization (PPO) are compared to linear proportional derivative (PD) controllers from classical control theory.
The RL agents reduced convection, measured by the Nusselt Number, by up to 33\% in moderately turbulent systems and 10\% in highly turbulent settings, clearly outperforming PD control in all settings.
The agents showed strong generalization performance across different initial conditions and to a significant extent, generalized to higher degrees of turbulence. The reward shaping improved sample efficiency and consistently stabilized the Nusselt Number to higher turbulence levels.

%We conclude that a single-agent DRL using PPO solves the presented RBC system for low turbulence, leaving little room for further improvement. However, plain PPO does not reliably find one-cell states for systems of high turbulence, leaving opportunities for further improvement. Future work should explore multi-agent RL in highly turbulent settings and model-based RL to address sample inefficiency and high computational costs.

\end{abstract}
\section{Introduction}
In recent years, Deep Reinforcement Learning (DRL) has demonstrated significant potential for control tasks in fluid dynamics, including turbulence suppression, optimal mixing, and drag reduction \cite{garnier_2021}. A key advantage of DRL is its ability to discover effective control strategies in highly nonlinear systems, for which conventional methods such as PID control often prove inadequate.

In this work, we study the effectiveness of model-free DRL for controlling natural convection dynamics governed by Rayleigh-Bénard Convection (RBC). Convection plays a crucial role in both natural and industrial processes, including oceanic circulation, cloud formation, stellar dynamics, and material processing \cite{beintema_2020,tang_1993}. In industry, methods to control convection are particularly important; for instance, in crystal growth processes, excessive convection can lead to instabilities that compromise material quality. The RBC system models a fluid confined between a heated lower plate and a cooler upper plate, where buoyancy-driven flows emerge throughout the fluid due to density differences originating from the bottom heating. Such convective flows become stronger and increasingly turbulent with higher temperature gradients and lower fluid viscosities\cite{pandey2018}. We specifically explore how effectively reinforcement learning, using Proximal Policy Optimization (PPO), can reduce convection under increasingly turbulent conditions, comparing its performance against conventional PD control.

A central challenge in RL applications in fluid dynamics is generalization. In RBC systems, even slight variations in initial temperature and velocity fields can yield significantly different flow structures. Thus, practical RL agents must exhibit robustness to variations in initial conditions, and ideally, the agents can generalize across different turbulence regimes without retraining.
Another critical challenge is sample efficiency. Training high-performing RL agents typically requires numerous rollouts, particularly when using model-free approaches combined with high-fidelity fluid simulations. Methods to alleviate the sample requirement not only improve training efficiency but also enhance the adaptability of RL agents to new scenarios.

Our main contributions explore the RBC control task using RL, addressing challenges as follows:
1) \textbf{Performance under increasing turbulence}:
We evaluate PPO's effectiveness in reducing convective motion across RBC systems with varying turbulence levels, comparing its performance quantitatively and qualitatively with PD control.
2) \textbf{Generalization ability}:
We train RL agents across different initial conditions and assess their generalization ability across unseen initial states and varying turbulence regimes.
3) \textbf{Training efficiency}: We introduce and demonstrate a reward-shaping technique that accelerates training and improves the agent's final performance.

\section{Related Work}
Controlling RBC to reduce or avoid convection has been researched for decades, first using methods from conventional control theory and later using data-driven approaches from Machine Learning.
Until recently, linear controllers like proportional (P) or proportional derivative (PD) control were widely used to stabilize the RBC system and reduce convection.
In the early 1990s, Tang and Bau \cite{tang_1993} established the theoretical foundation for stabilizing the RBC system through linear feedback control. Their work showed that starting from the no-motion state, a control input proportional to the midline temperature of the system could significantly delay the onset of convective heat transfer. 
Later, Howle demonstrated in a series of experiments that stabilizing RBC is possible in practical setups by placing a network of heaters in the lower boundary of the system \cite{howle1997}. Instead of relying on the midline temperature, he measured the vertically averaged density field using shadowgraph measurements for linear proportional control. In the 2000s, Remillieux et al. further investigated the suppression of convection in RBC using a PD controller, demonstrating its effectiveness in experimental and numerical simulation setups \cite{remillieux2007}.

In 2020, Beintema et al. \cite{beintema_2020} introduced Reinforcement Learning (RL) based control for RBC, outperforming earlier approaches built on PD control. Their approach stabilized RBC up to $\Ra=3*10^4$ and achieved a greater convection reduction, measured by the Nusselt Number, for Rayleigh Numbers above.
Vignon et al. \cite{vignon_2023} extended this work by proposing a multi-agent RL (MARL) approach to improve sample efficiency, which exploits translational invariance of individual heaters that act as agents, achieving 22.7\% reduction of the Nusselt number for $\Ra=10^4$ on a wide unrestricted domain in the horizontal direction that is closer to industrial applications.
Further work investigates extending the MARL approach to 3D-RBC \cite{vasanth2024} and including a positional encoding for the heaters \cite{jeon2024a}.

While \cite{vignon_2023} employs a highly scalable MARL framework in a moderately turbulent setting, this work examines the performance of more expressive single-agent RL agents in regimes of higher turbulence (i.e., larger Ra ). We adopt the same system parameters, actuation, and convection measures as \cite{vignon_2023}. Furthermore, we introduce reward shaping to increase sample efficiency and increase the practical feasibility of the agents by focusing specifically on the generalizability of the RL agents across initial conditions and turbulence regimes.

\section{Methodology} \label{sec:methods}
In this section, we present the RBC system, describe the simulation environment, and outline the control task and methods.

\subsection{Rayleigh-Bénard Convection} \label{sec:methods:rbc}
Rayleigh-Benard Convection (RBC) is a system that models conductive and convective heat transfer in a layer of fluid heated from below. The dynamics is governed by a partial differential equation (PDE) that is based on the incompressible Navier Stokes equations and can be found in \cite{pandey2018}, for instance.
The system's state in 2D can be fully described by the velocity vector field $\bm{u}=(u_x, u_y)$, the scalar temperature field $T$, and its initial and boundary conditions (see section \ref{sec:methods:sim}).
The Rayleigh Number Ra is a key system parameter and quantifies the strength of buoyancy-driven convection. It is proportional to the temperature gradient between the bottom and top layers, see \cite{pandey2018}. For increasing Ra, the flow becomes more unstable, resulting in higher degrees of convective turbulence \cite{vignon_2023}.

\begin{figure}[t]
     \centering
     \begin{subfigure}[b]{0.49\textwidth}
         \centering
         \includegraphics[width=0.88\textwidth]{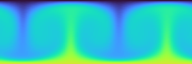}
         \caption{}
     \end{subfigure}
     \hfill
     \begin{subfigure}[b]{0.49\textwidth}
         \centering
         \includegraphics[width=0.88\textwidth]{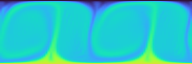}
         \caption{}
     \end{subfigure}
     \begin{subfigure}[b]{0.4\textwidth}
         \centering
         \includegraphics[width=\textwidth]{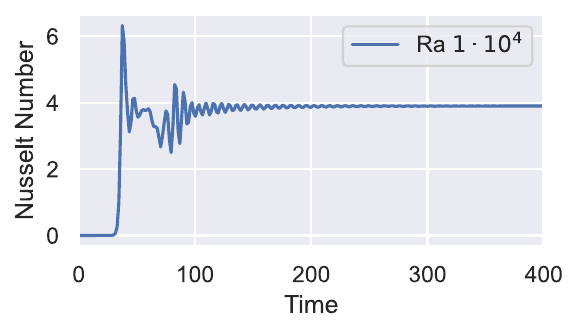}
         \caption{}
         \label{fig:baseline_episode_nusselt}
     \end{subfigure}
     \hfill
     \begin{subfigure}[b]{0.44\textwidth}
         \centering
         \includegraphics[width=\textwidth]{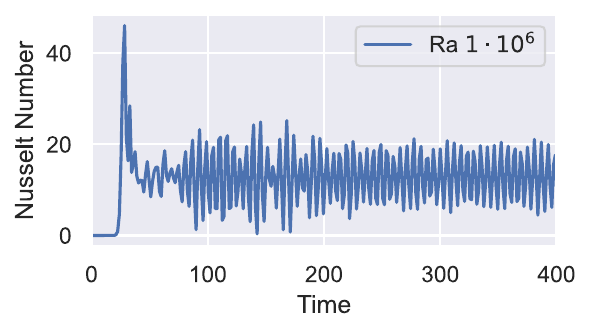}
         \caption{}
     \end{subfigure}
     \caption{Figure (a) shows an example of a temperature field for the uncontrolled system at $\Ra=10^4$ and the line in (c) shows Nu for the full roll-out. Figures (b) and (c) show the same for an uncontrolled system at $\Ra=10^6$.}
    \label{fig:baseline}
\end{figure}

The strength of convection in the RBC system is measured by the local convective heat flux given by equation 
\begin{equation} \label{eq:convective}
    q(x,y,t) = u_y(x,y,t) \theta(x,y,t) \,, \text{ with } \theta(x,y,t) = T(x,y,t) - \langle T \rangle_{x, y}\,,
\end{equation}
where $\theta$ denotes the temperature fluctuations around the mean temperature over the domain $\langle T \rangle_{x,y}$.
% is the temperature field as a deviation from the spatial average temperature $\langle T \rangle_{x,y}$. 
This leads to the Nusselt Number Nu, which measures the amount of convection and is defined as the ratio of convective to conductive heat transfer. Following previous studies \cite{beintema_2020,vignon_2023,markmann2024}, we define Nu as:\footnote{$T_b, T_t$: temperature at bottom (heating) and top, resp. $H$: distance between top and lower boundary. $\kappa:$ thermal diffusivity.}
\begin{equation}\label{eq:nusselt}
    \Nu (t) = \frac{\langle q(x, y, t) \rangle_{x, y}}{\kappa (T_b - T_t) / H}\,.
\end{equation}
Initially, the heat originating from the bottom is only transferred by conduction, and the temperature varies linearly in the vertical direction. When the Rayleigh Number exceeds the critical threshold $\Ra_c=1708$, convection starts, and the fluid organizes into Bénard cells (see Figure \ref{fig:baseline}). Since in this state, heat is transferred more by convection than conduction, $\Nu > 1$. As Ra increases further, the fluid flow transitions from structured to turbulent convection with increasingly chaotic behavior.

\subsubsection{Simulation Environment} \label{sec:methods:sim}
We use a Direct Numerical Simulation (DNS) to solve the RBC system based on the open-source framework Shenfun \cite{mortensen_joss}. The governing equations are numerically solved using the spectral Galerkin method on a 2D rectangular spatial domain.\footnote{Spatial Dimensions: horizontal $x \in [0,2\pi]$, vertical $y \in [-1,1]$, $H=2$ , discretized to $96 \times 64$ uniform grid points. Our code repository provides further parameters and equations: \url{https://github.com/HammerLabML/RBC-Control-SARL}} Our simulation setup follows previous studies \cite{beintema_2020,vignon_2023,markmann2024}, where Boundary Conditions (BC) include no-slip walls at the bottom and top and periodic BCs in the horizontal direction. The initial condition is given by small perturbations added to the conductive equilibrium. The system is evolved with a time step of $\Delta t = 0.025$.

\subsection{Control of RBC} \label{sec:methods:rbc_control}
Following previous studies \cite{howle1997,beintema_2020,vignon_2023}, we focus on the relevant control task of reducing Nu in the convective system by applying small temperature fluctuations at the lower boundary. The lower wall is discretized into $N=12$ heating segments, each receiving independent control inputs $a_i$ for $i=1,\dots N$. To maintain a constant Ra, the applied control is centered and the values are constrained to $[-C,C]$, with $C=0.75$, using the following consecutive transformations from \cite{beintema_2020,vignon_2023}:
\begin{equation} \label{eq:transform}
\hat T'_i = a_i - \frac{\sum_{i=1}^{N} a_i}{N}\,, \quad \hat T_i = \frac{\hat T'_i C}{\max(1, |T'|)} \quad . % \quad \hat T_i = \text{smooth}(\hat T_i) 
\end{equation}
The transformed control inputs $\hat T_i$ are then mapped to the horizontal spatial dimension with additional smoothing between the heating segments\cite{vignon_2023}. This ensures stability of the numerical simulation.
%Figure \ref{fig:control_example} illustrates a sample control input applied to the RBC system.
%
%\begin{figure}
%    \centering
%     \begin{subfigure}[b]{0.45\textwidth}
%         \centering
%         \includegraphics[width=\textwidth]{figure/methods/state.png}
%         \caption{State with control}
%     \end{subfigure}
%     \hfill
%     \begin{subfigure}[b]{0.45\textwidth}
%         \centering
%         \includesvg[width=\textwidth]{figure/methods/action.svg}
%         \caption{Temperature Fluctuation}
%     \end{subfigure}
%    \caption{Control Input example}
%    \label{fig:control_example}
%\end{figure}

%\begin{table}
%    \caption{Simulation parameters for Rayleigh-Bénard Convection.}\label{tab:sim_params}
%    \centering
%    \begin{tabular}{|l|c|}
%        \hline
%        \multicolumn{2}{|c|}{Simulation} \\
%        \hline
%        Parameter &  Value\\
%        \hline
%        Domain Size             &  $2\pi \times 2$    \\
%        Mesh Grid               &  $64 \times 96$     \\
%        Ra                      &  $\{ 1e4, 1e5, 1e6, 5e6\}$     \\
%        Pr                      &  0.7     \\
%        ($T_t$, $T_b$)          &  $(1,2)$     \\
%        $\Delta t$              &  0.025     \\
%        Baseline Length         &  400  \\
%        \hline
%    \end{tabular}
%\end{table}

\subsubsection{Linear Control} \label{sec:methods:linear}
A linear proportional-derivative controller (PD) serves as a baseline for reducing convection in the RBC system \cite{remillieux2007,beintema_2020}. The temperature fluctuations at the lower boundary are computed via
\begin{flalign} \label{eq:pdcontrol}
a(x, t) = k_p E(x,t) + k_d  E'(x,t) \,,
\end{flalign}
where $k_p$ and $k_d$ are the proportional and derivative gains, $E(x, t)$ represents the distance from a desired state, which we define as the deviation from the no-motion equilibrium state $u_y^*=0$ as in \cite{beintema_2020}:
\begin{flalign} \label{eq:pdcontrol:err}
E(x,t) = \langle u_y(x,y,t) \rangle_y - u_y^{*}=\langle u_y(x,y,t) \rangle_y \hspace{0.2cm} \quad .
\end{flalign}
The PD control strategy is to oppose convection by applying heat to cold, downflowing regions between Bénard cells while reducing heat elsewhere. We use the controller gains of $k_p$=970 and $k_d$=2000. The resulting control input $a(x, t)$ is discretized in $N=12$ segments by averaging corresponding grid points and transformed via Eq.~\eqref{eq:transform}.

\subsubsection{Reinforcement Learning} \label{sec:methods:rl}
Reinforcement Learning (RL) is an approach of learning by trial and error, assuming that rewarding an agent (with a quantity $R$) for desired behavior (an action $a$) in a state of the environment $s$, leads to reinforcing this behavior in the future\cite{suttonbarto}. This setting is similar to conventional feedback control, but RL can discover complex control policies through expressive neural networks.
The agent follows a policy  $\pi(a|s)$, which maps states to action probabilities. The objective of RL is to find a policy $\pi^*$ that maximizes the expected sum of reward,
$\max_{\pi} \mathbb{E} \left [ \sum_{t=0}^{\infty} \gamma^t R(s_t,a_t,s_{t+1}) \right ]$,
where actions $a_t$ are chosen according to $\pi$ and $0 \leq \gamma \leq 1$ discounts future rewards. The state transitions are determined by a Markov Decision Process (MDP), in our case given by the underlying deterministic numerical simulation.

% \begin{itemize}
%     \item $S$ is the set of states,
%     \item $A$ is the set of possible actions,
%     \item $T(s'|s,a)$ is the transition function defining the probability of reaching state  $s'$ from state $s$ after taking action $a$,
%     \item $R(s',a,s)$ is the reward function,
%     \item $\gamma \in [0,1]$ is the discount factor,
%     \item $H$ is the time horizon.
% \end{itemize}

For the observations, we assume probe sensors on a $8 \times 48$ grid over the spatial domain that measure the local temperature and velocity field\cite{vignon_2023, beintema_2020}. The measurements are flattened into a vector of size $1152$ as input to the agent.
The action space consists of temperature fluctuations applied at the heating segments at the lower boundary $(a_1, a_2,\dots,a_N)$, where $a_i \in [-1,1]$, which are transformed via Eq.~\eqref{eq:transform}.

The agent's objective is to minimize convective heat transfer as measured by the Nusselt Number Nu from Eq.~\eqref{eq:nusselt}, which is reflected by the reward:
\begin{flalign}
    %r_t = 1 - \frac{Nu(t)}{Nu_{Base}(Ra)} \, ,
    R(s_t) = 1 - \frac{\text{Nu}(s_t)}{\text{Nu}_{\mathrm{Base}}(\Ra)} \, ,
\end{flalign}
where $\text{Nu}_{\text{Base}}(\Ra)$ is the maximum Nu occurring in the uncontrolled system at the given Ra, so that approximately $R(s_t) \in [0,1]$.

The control policy is trained using PPO\cite{schulman2017}, a model-free, actor-critic policy optimization technique, which improves trained agents' robustness by employing a clipped objective function that avoids large policy updates.
Due to its balance between performance and stability, PPO is widely used across various application domains, including control of RBC \cite{beintema_2020, vignon_2023}.

\subsection{Reward Shaping} \label{sec:methods:rew_shaping}
Our preliminary results indicated that simple strategies, such as heating between cells as PD control does, can also significantly reduce Nu, but do not stabilize the flows. Observations in \cite{vignon_2023} for $\Ra=10^4$ and our first results identified cell merging as an effective strategy for reducing Nu and stabilizing the flow.
However, PPO agents were often trapped in a similar strategy as the PD control, while training agents capable of merging cells required significant training effort. To address this, we experiment with reward shaping to incentivize cell merging.

We detected potential cell locations ${c_i}$ by finding positive local maxima of the vertical velocity measurement $u_y(x,0,t)$ at the vertical midline of the domain using simple numerical peak finding.\footnote{We used find\_peaks from scipy.signal with height=0. The robustness of the peak finding in turbulent flows may be improved by tuning the parameters of the peak finding algorithm.}
To quantify the degree of merging, we compute the maximum pairwise cell distance:
\begin{equation}
    \text{celldist} = \max \left(\{ \min(|c_i - c_j|, 2\pi - |c_i - c_j|) \, | \, 1 \leq i < j \leq K \}\right)\,,
\end{equation}
where the $\min$ function ensures the correct distance between cells $i$ and $j$ in the periodic domain $x \in [0, 2\pi]$. If at most one cell exists, we set $\text{celldist}=0$.
%\footnote{For our system parameters, we exclusively deal with two-cell initial states.}.
To promote cell merging while maintaining a low Nusselt number, we modify the reward function:
\begin{equation} \label{eq:reward_shaping_reward}
    r_t = (1-\alpha) \left( 1 - \frac{\Nu(t)}{\Nu_{\text{Base}}(\Ra)} \right)  + \alpha \left (1 - \frac{\text{celldist}(t)}{\pi} \right) \,,
\end{equation}
 where $\alpha \in [0,1]$ balances the cell distance and the Nusselt number in the reward. The quantity $(1-\text{celldist}(t)/\pi)$ ranges from $0$, when the cells are maximally separated,\footnote{Note that $\pi$ is the maximum possible distance on the periodic domain $x \in [0, 2\pi]$.} to $1$, in case of a single merged cell.

\section{Experiments and Results}
We conduct three experiments to evaluate the effectiveness of DRL in comparison with linear control for increasing turbulence in the RBC system.
First, we explain how the environment rollouts and training of the DRL method work. In experiment 1, we compare DRL and linear control in reducing the system's convection. Next, we demonstrate the agents' generalizability to other turbulence levels in experiment 2 and introduce the reward-shaping technique to the DRL approach in experiment 3.

\subsection{Episode Rollouts and PPO training}
We trained and evaluated the methods on environment rollouts using the DNS, which is wrapped in a Gym environment \cite{towers2024gymnasium}.\footnote{Observation Grid: $8 \times 48$, Heating segments $N$: $12$, Action limit $C$: $0.75$. Action duration $1.5$, Episode length: $300$, i.e. 200 actions per episode.}
In the next experiments, we evaluated the DRL and linear control at different turbulence levels, i.e. $\Ra \in \{1e4,1e5,1e6,5e6\}$, keeping the other parameters fixed.
Fig \ref{fig:baseline} illustrates two uncontrolled simulations of RBC over 400 timesteps at different Ra, starting from the no-motion state and transitioning to the convection phase. At lower turbulence ($\Ra=1e4$), the system eventually converges to a stable state with Nu of around $\sim$3.9.
In contrast, higher turbulence leads to a periodic behavior with typically two convection cells (Fig.~\ref{fig:exp1:t0}). Occasionally, the system converges to a single cell, resulting in lower Nu in the attractor.

As in \cite{vignon_2023}, the agents control the system only in its convective phase, starting 400 time steps after the initial condition, ensuring that the rollouts begin after convection has been established rather than from the no-motion state.
We created 35 checkpoints of different convective states per Ra, resulting from 35 initial conditions of the system.
To evaluate whether the methods generalize well over different initial conditions, we organized the checkpoints into train, validation, and test sets of size 20, 5, and 10, respectively.

%\begin{table}
%    \caption{Environment parameters for Rayleigh-Bénard Convection rollouts.}\label{tab:exp:env_params}
%    \centering
%    \begin{tabular}{|l|c|}
%        \hline
%        \multicolumn{2}{|c|}{RL environment} \\
%        \hline
%        Parameter &  Value\\
%        \hline
%        Observation Grid            & $8 \times 48$      \\
%        Episode Length              & 300       \\
%        Action scaling factor       & 0.75      \\
%        Action Duration             & 1.5      \\
%        Number of Control Segments  & 12      \\
%        \hline
%    \end{tabular}
%\end{table}

The PPO-based agents are trained for 400,000 action steps, corresponding to 2000 single rollouts of the RBC system.\footnote{PPO implementation from Stable-Baselines-3\cite{sb3}, $20$ parallel environments yielding 4,000 samples per iteration, $\gamma =0.99$, entropy $\beta = 0.01$, two-layer (64 hidden units) neural networks for both the actor and critic.}
To avoid overfitting, we continuously validate the agent's performance on the validation checkpoints, storing the agent with the highest mean return.
For the best agent, we recorded the reductions in Nu for the 10 test checkpoints.

%\begin{table}
%    \centering
%    \caption{PPO training parameters}\label{tab:exp:ppo}
%    \begin{tabular}{|l|c|}
%        \hline
%        Hyperparameter &  Value \\
%        \hline
%        Parallel Environments       & 20        \\
%        Total Environment Steps     & 400,000   \\
%        Rollouts Per Iteration      & 1 x 20    \\
%        Batch Size                  & 4000      \\
%        Discount                    & 0.99      \\
%        Entropy Coefficient         & 0.01      \\
%        Learning Rate               & 0.001     \\
%        Hidden Layers               & 2         \\
%        Number of Neurons Per Layer \hspace{0.5cm} & 64        \\
%        Activation Function         & ReLU      \\
%        \hline
%    \end{tabular}
%\end{table}

%\begin{figure}
%    \centering
%    \includesvg[width=0.7\textwidth]{figure/data/train.svg}
%    \caption{Training of DRL agents. mention over fitting for higher turbulence maybe}
%    \label{fig:exp:train}
%\end{figure}

\subsection{Experiment 1: Nusselt Number Reduction} \label{sec:experiment1}
\begin{figure}[t]
    \begin{minipage}{0.48\textwidth}
        \footnotesize
        \begin{tabular}{|r||r|r|r|}
            \hline
            \multicolumn{4}{|c|}{\textbf{Mean Nusselt Number}} \\
            \hline
            \textbf{Ra} & \textbf{Baseline} & \textbf{PD} & \textbf{PPO}  \\
            \hline
            $10^4$      & 3.9  \scriptsize $\pm 0.00$       & 3.1  \scriptsize $\pm 0.02$    & \textbf{2.6}  \scriptsize $\pm 0.03$     \\
            $10^5$      & 6.9  \scriptsize $\pm 0.01$       & 6.9   \scriptsize $\pm 0.02$    & \textbf{5.9}  \scriptsize $\pm 0.18$     \\
            $10^6$      & 11.6 \scriptsize $\pm 0.38$       & 12.5  \scriptsize $\pm 0.48$    & \textbf{11.2} \scriptsize $\pm 0.35$     \\
            $5 \cdot 10^6$      & 19.8 \scriptsize $\pm 0.16$       & 23.2  \scriptsize $\pm 0.14$    & \textbf{17.5} \scriptsize $\pm 0.21$     \\
            \hline
        \end{tabular}
    \end{minipage}
    \hfill
    \begin{minipage}{0.44\textwidth}
        \includegraphics[width=\textwidth]{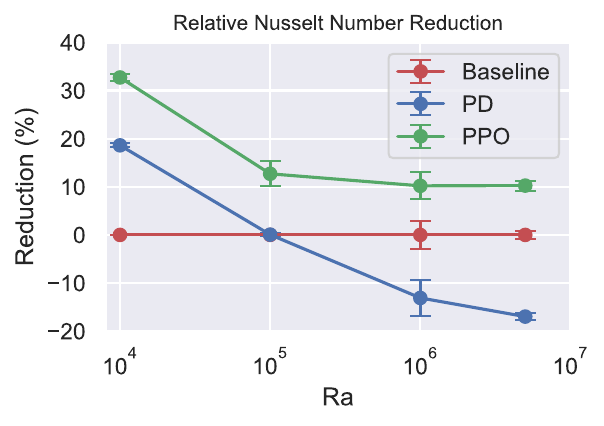}
    \end{minipage}
    \caption{Left: Episode-averaged Nusselt number computed for the 20 test initial conditions for the uncontrolled system (baseline), PD- and PPO control for different Ra. Right: Relative reduction with respect to the baseline.}
    \label{fig:exp1}
\end{figure}

We evaluated the effectiveness of the DRL agent and PD control in reducing Nu across all turbulence levels.
Figure \ref{fig:exp1} presents the episode-averaged Nu for each Ra and the relative reduction with respect to the uncontrolled baseline. 

\begin{figure}[t]
    \centering
    \begin{subfigure}[b]{0.52\linewidth}
        \centering
        \includegraphics[width=0.85\textwidth]{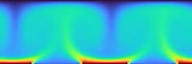}%
        \includegraphics[width=0.15\textwidth]{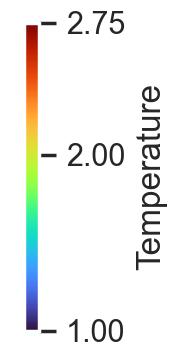}
        \caption{$t=30$}
    \end{subfigure}
    \hfill
    \begin{subfigure}[b]{0.42\linewidth}
        \centering
        \includegraphics[width=\textwidth]{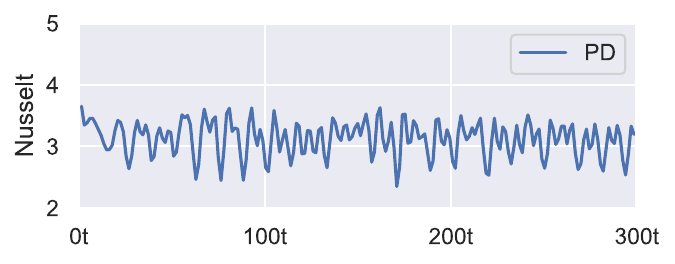}
        \caption{Nusselt Reduction of rollout}
    \end{subfigure}
    \caption{Example of linear PD control for an $\Ra=10^4$ rollout of the system.}
    \label{fig:exp1:pd}
\end{figure}

PD control worked well for the least turbulent case ($\Ra=10^4$), reducing the Nusselt number by 19\%. For $\Ra = 10^6$ and $\Ra = 5*10^6$, PD control resulted in states larger Nu than the uncontrolled baseline.
Fig.~\ref{fig:exp1:pd} visualizes the PD control's strategy at $\Ra=10^4$, only applying heat between cells.
This keeps the system in a two-cell setup, lowering Nu to around $\sim$3.1 with some variation during the rollout.

\begin{figure}[t]
    \centering
    \begin{subfigure}[b]{0.45\linewidth}
        \centering
        \includegraphics[width=\textwidth]{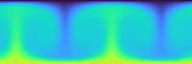}
        \caption{$t=0$}
        \label{fig:exp1:t0}
    \end{subfigure}
    \hfill
    \begin{subfigure}[b]{0.45\linewidth}
        \centering
        \includegraphics[width=\textwidth]{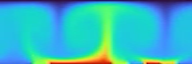}
        \caption{$t=12$}
        \label{fig:exp1:t12}
    \end{subfigure}
    \begin{subfigure}[b]{0.45\linewidth}
        \centering
        \includegraphics[width=\textwidth]{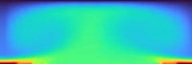}
        \caption{$t=60$}
        \label{fig:exp1:t60}
    \end{subfigure}
    \hfill
    \begin{subfigure}[b]{0.45\linewidth}
        \centering
        \includegraphics[width=\textwidth]{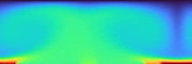}
        \caption{$t=210$}
        \label{fig:exp1:t210}
    \end{subfigure}
    \begin{subfigure}[b]{0.45\linewidth}
        \centering
        \includegraphics[width=\textwidth]{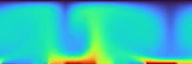}
        \caption{$t=270$}
        \label{fig:exp1:t270}
    \end{subfigure}
    \hfill
    \begin{subfigure}[b]{0.45\linewidth}
        \centering
        \includegraphics[width=\textwidth]{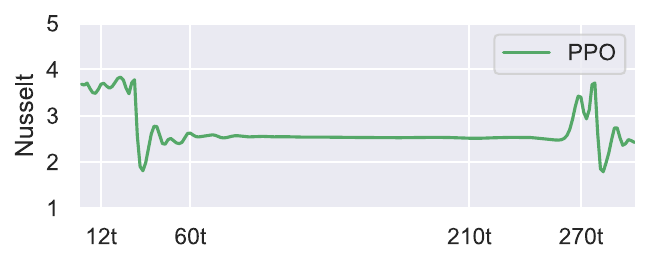}
        \caption{Nusselt Reduction of rollout}
        \label{fig:exp1:ppo_1e4:nu}
    \end{subfigure}
    \caption{Example of PPO control for an $\Ra = 10^4$ test set rollout of the system.}
    \label{fig:exp1:ppo_1e4}
\end{figure}

In contrast, DRL agent consistently achieved much better results across all Ra, with a 33\% reduction for $\Ra=10^4$, a 15\% reduction for $\Ra=10^5$, and around 10\% for $\Ra > 10^5$.
Fig.~\ref{fig:exp1:ppo_1e4} illustrates the DRL control strategy in a low-turbulence system ($\Ra=10^4$). Within 60 timesteps, the agent actively merged the two cells into a single cell (Fig.~\ref{fig:exp1:t60}) to reduce the total convection and stabilize Nu to a value of $2.6$.
Afterward, the agent maximized the single cell's width (Fig.\ref{fig:exp1:t210}), sometimes causing a return to a two-cell state (Fig.~\ref{fig:exp1:t270}). Since this behavior did not worsen the mean return, as shown in Figure \ref{fig:exp1:ppo_1e4:nu}, this was not penalized.
The control strategy differed at higher turbulence levels $\Ra>10^4$: Instead of driving the system to one-cell states, the agent resorted to a slight adaptation of the PD control's strategy, which still effectively reduced Nu. \footnote{Further results and videos of the agents are available on \url{https://github.com/HammerLabML/RBC-Control-SARL}}
%\begin{figure}
%    \centering
%    \begin{subfigure}[b]{0.45\linewidth}
%        \centering
%        \includegraphics[width=\textwidth]{figure/exp1/2_t0.png}
%        \caption{$t=0$}
%    \end{subfigure}
%    \hfill
%    \begin{subfigure}[b]{0.45\linewidth}
%        \centering
%        \includegraphics[width=\textwidth]{figure/exp1/2_t30.png}
%        \caption{$t=30$}
%    \end{subfigure}
%    \begin{subfigure}[b]{0.45\linewidth}
%        \centering
%        \includegraphics[width=\textwidth]{figure/exp1/2_t270.png}
%        \caption{$t=270$}
%    \end{subfigure}
%    \hfill
%    \begin{subfigure}[b]{0.45\linewidth}
%        \centering
%        \includesvg[width=\textwidth]{figure/exp1/nusselt_example_2.svg}
%        \caption{Nusselt Reduction of rollout}
%    \end{subfigure}
%    \caption{Example of PPO control for a Ra $1 \cdot 10^6$ based rollout of the system.}
%    \label{fig:exp1:ppo_1e6}
%\end{figure}

\subsection{Experiment 2: Generalization across turbulence levels} \label{sec:experiment2}
\begin{figure}[t]
     \centering
     \begin{subfigure}[b]{0.44\textwidth}
         \centering
         \includegraphics[width=\textwidth]{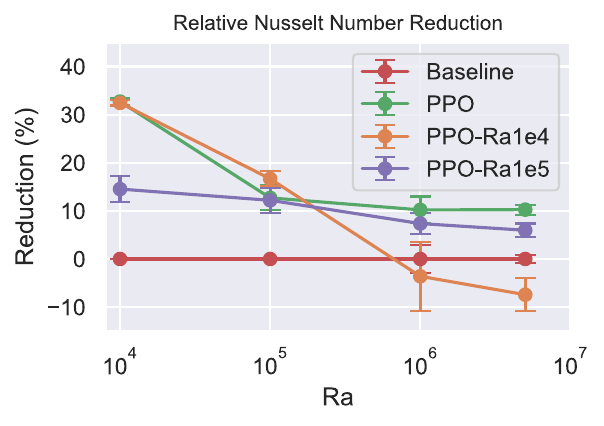}
         \caption{}
         \label{fig:exp2}
     \end{subfigure}
     \hfill
     \begin{subfigure}[b]{0.44\textwidth}
         \centering
         \includegraphics[width=\textwidth]{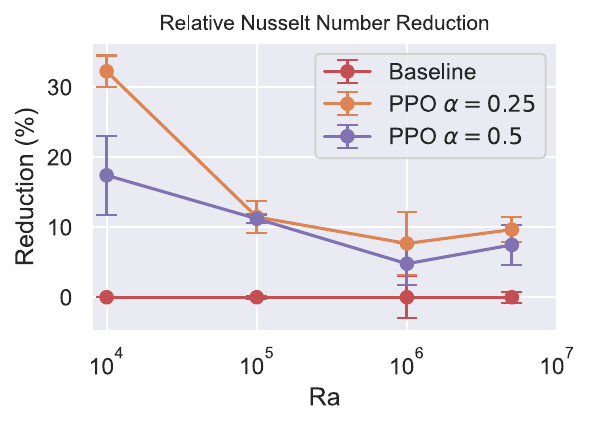}
         \caption{}
         \label{fig:exp3:nusselt_reduction_percentage_RS}
     \end{subfigure}
     \caption{(a) Reduction of Nu on the 20 test initial conditions in the generalization task of Sec.~\ref{sec:experiment2} and (b) when training with reward shaping (Sec.~\ref{sec:experiment3}). The green line in (a) is the result from Sec.~\ref{sec:experiment1} and acts as a baseline here.}
     \label{fig:exp2_exp3_nusselt_reduction}
    %\label{fig:nusselt_reduction_percentage_RS}
\end{figure}

Given the hierarchical structure of turbulence,
agents trained in low-turbulence systems may learn control patterns
effective
in higher-turbulence environments.
We evaluated agents trained at $\Ra=10^4$ and $\Ra=10^5$ in their ability to reduce convection across different Ra. 
Fig.~\ref{fig:exp2} shows the relative reduction of Nu compared to the uncontrolled baseline and the PPO baseline of the previous experiment.
PPO-Ra1e4 performed even better on $\Ra=10^5$ than the PPO baseline, successfully carrying over the cell merging strategy, but failing for higher Ra.
PPO-Ra1e5 achieved Nu reduction across all Ra, however, its performance was worse than the PPO baseline and never merged cells.
A possible explanation of these different behaviors is that the agent trained for $\Ra=10^4$ acquired the cell merging strategy due to the more stable flows in the system, while flows are more chaotic for higher Ra, making learning of adequate control more difficult.

\subsection{Experiment 3: Reward Shaping} \label{sec:experiment3}
\begin{figure}[t]
     \centering
     \begin{subfigure}[b]{0.32\textwidth}
         \centering
         \includegraphics[width=\textwidth]{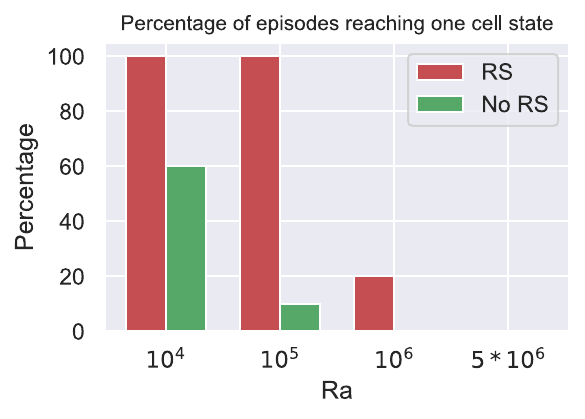}
         \caption{}
         \label{fig:exp3:percentage_episodes_one_cell}
     \end{subfigure}
     \begin{subfigure}[b]{0.32\textwidth}
         \centering
         \includegraphics[width=\textwidth]{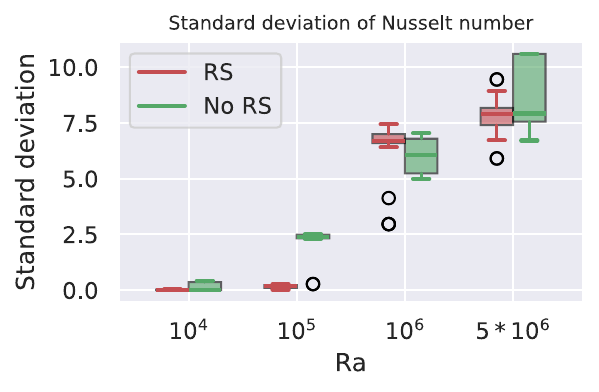}
         \caption{}
         \label{fig:exp3:reduction_nusselt_std}
     \end{subfigure}
     \hfill
     \begin{subfigure}[b]{0.32\textwidth}
         \centering
         \includegraphics[width=\textwidth]{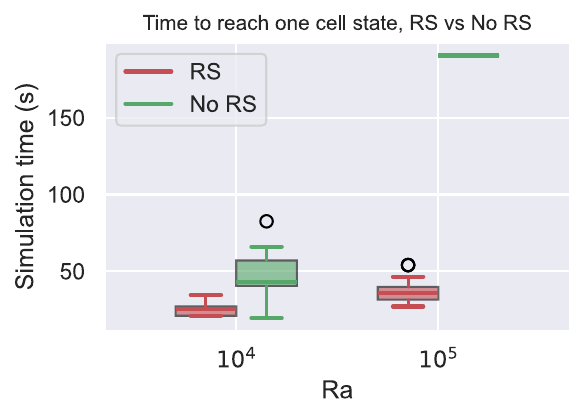}
         \caption{}
         \label{fig:exp3:time_eps_onecell}
     \end{subfigure}
     \caption{The effect of reward shaping in several statistics.}
\end{figure}

We evaluated the impact of introducing the reward function \eqref{eq:reward_shaping_reward} with balancing values of $\alpha=0.25$ and $\alpha=0.5$ for each Ra.
We refer to agents trained with reward shaping as RS-agents and those trained without it as No-RS-agents.
Fig.~\ref{fig:exp3:nusselt_reduction_percentage_RS} shows that the Nusselt number reduction is comparable to that of No-RS-agents (Fig.~\ref{fig:exp1}).
However, RS-agents demonstrated a significantly higher success rate in merging convection cells, as shown in Fig.~\ref{fig:exp3:percentage_episodes_one_cell}: Cells were always merged for $\Ra=10^4$ and $\Ra=10^5$, and even for $\Ra=10^6$, merging remained possible. In contrast, No-RS-agents showed significantly lower cell merging.
%We note that the No-RS-agents also achieved 100\% cell merging for $\Ra=10^4$, but it only shows 60\% due to the break ups of the single cell as shown in Fig.~\ref{fig:exp1:t270}. 

The cell merging strategy of RS-agents was reflected in the variation of Nu over time during the final 40 steps of the rollouts, as illustrated in
Fig.~\ref{fig:exp3:reduction_nusselt_std}:
RS-agents achieved a near-complete reduction in Nu variation for $Ra=10^4$ and $10^5$, resulting from the more stable one-cell flows, similar to the behavior of the Nusselt number in Fig.~\ref{fig:baseline_episode_nusselt}.
Additionally, for $\Ra=10^6$, episodes that reached a one-cell state had considerably less variation in Nu, which are shown as outliers in the box plot.

RS-agents merged cells early in the episode and faster than the No-RS agent, as shown in Fig.~\ref{fig:exp3:time_eps_onecell}.
Notably, we found that RS-agents trained on $\Ra=10^5$ consistently merged cells even at $\Ra=10^6$, yielding a significant reduction in Nusselt number variation.

\section{Discussion}
Our results demonstrate that Single-Agent RL performs effectively across a variety of turbulence levels, significantly outperforming PD control through the discovery of non-trivial control strategies. The observed 33\% reduction in the Nusselt number for $\Ra=10^4$ is higher than what was achieved in \cite{vignon_2023}, likely due to the greater expressiveness of Single-Agent RL. Hence, Single-Agent RL strategies should be regarded as an upper bound on achievable performance. This allows for a better interpretation of the performance of highly scalable yet less expressive methods, such as in \cite{vignon_2023}. Scalability is a limitation of the Single Agent setup: although we have demonstrated that training high-performing agents in a 2D domain can be achieved, it is unclear whether it remains feasible in 3D settings \cite{vasanth2024}.
However, reward shaping significantly alleviated the sample complexity, enabling agents to stabilize the flows quickly and consistently up to much higher Rayleigh numbers.

Our trained agents showed strong generalization performance: All agents trained for $\Ra=10^4$ and $\Ra=10^5$ employed their non-trivial strategy successfully on various test initial conditions. Additionally, they generalized to higher Rayleigh numbers: Agents trained for $\Ra=10^4$ (without reward shaping) were able to merge convection cells consistently for $\Ra=10^5$. Agents trained for $\Ra=10^5$ (with reward shaping) merged cells for $\Ra=10^6$. While the one-cell states for higher Ra were less stable than those achieved by agents trained directly for those settings, minimal fine-tuning may suffice to reduce all fluctuations.
We hypothesize that the increased turbulence at higher Ra makes agent training more challenging, preventing cell-merging strategies from always being discovered successfully. 
It may be possible that different actuation parameters may achieve stable flows more easily for higher Ra. At the same time, additional constraints arising from practical settings, such as limitations on heater temperatures and actuation duration, must be considered.

\section{Conclusion}
In this work, we trained RL agents to control Rayleigh-Bénard Convection. The agents discovered non-trivial strategies and exhibited strong generalization performance, making them viable for practical situations. The reward shaping and the ability to generalize across Rayleigh numbers highlight the potential for sample-efficient learning.

In future work, we aim to improve sample efficiency further using a model-based RL framework, integrating neural operator surrogate models for RBC, as explored in our preliminary work in \cite{straat_2025}. Additionally, we will investigate our approaches in the more realistic 3D setting.

\begin{credits}
\subsubsection{\ackname} 
\footnotesize %
The authors acknowledge financial support by the project ”SAIL: SustAInable Life-cycle of Intelligent Socio-Technical Systems” (Grant ID NW21-059A and NW21-059D), which is funded by the program ”Netzwerke 2021” of the Ministry of Culture and Science of the State of North Rhine Westphalia, Germany.
\end{credits}
\bibliographystyle{splncs04}
\bibliography{bibliography}

\end{document}